\documentclass[lettersize,journal]{IEEEtran}
\usepackage{amsmath,amsfonts}
\usepackage{algorithmic}
\usepackage{algorithm}
\usepackage{array}
\usepackage[caption=false,font=normalsize,labelfont=sf,textfont=sf]{subfig}
\usepackage{textcomp}
\usepackage{stfloats}
\usepackage{url}
\usepackage{verbatim}
\usepackage{graphicx}
\usepackage{cite}
\usepackage{CJKutf8}

\usepackage{hyperref}
\usepackage{mathrsfs}

\usepackage{textcomp}
\usepackage{stfloats}
\usepackage{url}
\usepackage{verbatim}
\usepackage{graphicx}
\usepackage{cite}
\usepackage{algorithm}
\usepackage{algorithmic, mathrsfs, bm}
\usepackage{amsmath}
\usepackage{amssymb}
\usepackage{mathtools}
\usepackage{amsthm}
\usepackage{multirow}
\usepackage{hyperref}
\usepackage{cleveref}
\usepackage{booktabs}

\newtheorem{mypro}{Problem}

\begin{document}

\title{Modeling the Popularity of Events on Web by Sparsity and Mutual-Excitation Guided Graph Neural Network}

\author{Jiaxin Deng, Linlin Jia, Junbiao Pang and Qingming Huang %Anonymous Author(s)

\IEEEcompsocitemizethanks{
\IEEEcompsocthanksitem J. Deng, L. Jia and J. Pang are with the Faculty of Information Technology, Beijing University of Technology, Beijing 100124, China (email: \mbox{junbiao\_pang@bjut.edu.cn}).

\IEEEcompsocthanksitem  Q. Huang is with the University of Chinese Academy of Sciences, Chinese Academy of Sciences (CAS), Beijing 100049, China, and the Institute of Computing Technology, CAS, Beijing
100190, China (email: qmhuang@ucas.ac.cn).
 }
}



\maketitle

\begin{abstract}
The content of a webpage described or posted an event in the cyberspace inevitably reflects viewpoints, values and trends of the physical society. Mapping an event on web to the popularity score plays a pivot role to sense the social trends from the cyberspace. However, the complex semantic correspondence between texts and images, as well as the implicit text-image-popularity mapping mechanics pose a significant challenge to this non-trivial task. In this paper, we address this problem from a viewpoint of understanding the interpretable mapping mechanics. Concretely, we organize the keywords from different events into an unified graph. The unified graph facilitates to model the popularity of events via two-level mappings, \emph{i.e.}, the self excitation and the mutual excitation. The self-excitation assumes that each keyword forms the popularity while the mutual-excitation models that two keywords would excite each other to determine the popularity of an event. Specifically, we use Graph Neural Network (GNN) as the backbone to model the self-excitation, the mutual excitation and the context of images into a sparse and deep factor model. Besides, to our best knowledge, we release a challenge web event dataset for the popularity prediction task. The experimental results on three public datasets demonstrate that our method achieves significant improvements and outperforms the state-of-the-art methods. Dataset is publicly available at: \url{https://github.com/pangjunbiao/Hot-events-dataset}.
\end{abstract}


\section{Introduction}

\begin{figure}[t!]
    \centering
    \includegraphics[scale=0.27]{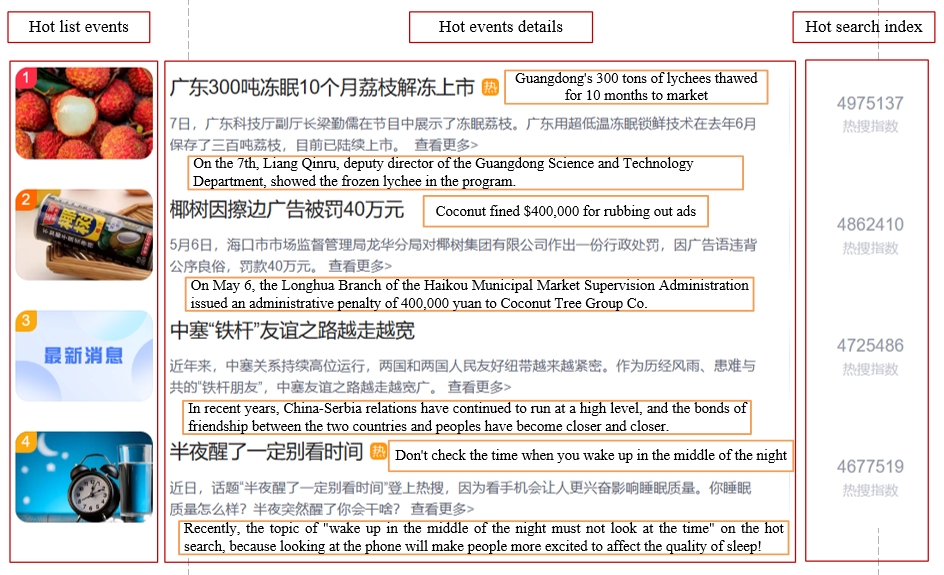}
    \caption{Some events and their popularity scores released by our dataset originally from the Baidu popularity list.}
    \label{fig:examplebaidu}
\end{figure}

Internet is a convenient and efficient method for the public to obtain information in time. Online media (\emph{e.g.}, blogs, online news) have greatly expanded the channels of information dissemination, reporting the current social events. With the booming of social posts, the issue of significant information overload for ordinary users has arisen, and it becomes even more serious with the emergence of content aggregation and distribution platforms. In order to alleviate this issue, the popularity-based content display is proposed~\cite{wang2018learning} to rank content by their popularity scores, measuring the quality of the interactions among users and content, as well as the browsing or view time of an event (as illustrated in Fig.~\ref{fig:examplebaidu}). 

Consequently, the free, and open nature of online media~\cite{manning1999foundations} makes it difficult to understand why an event has a great popularity score to the public~\cite{joo2017impact}. Therefore, a non-trivial task is to predict the popularity score for a newly emerging event on the web. It not only helps users find the popular content ahead of time, but also supports the content distribution platform~\cite{figueiredo2011tube} by understanding which content is intensively interesting to the public, and even which word could deeply resonance to the public.

Principally, the subliminal meaning of text and images used in an event~\cite{van2008persuasiveness} is the key factor for measure popularity. However, mapping both text and images from an event to the popularity scores suffers from the following problems:
\begin{itemize}
\item \textbf{Interpretable mapping mechanism from text to popularity:} ~\cite{martin2016exploring}~\cite{wang2018learning}~\cite{duan2023multi} predicted the popularity
of user generated context (UGC) from the perspectives of feature engineering. Unfortunately, they ignored the interpretability of the results. One of our observations is that sparse keywords (see Table~\ref{fig:examplebaidu}) could help a deep model to interpret the popularity mapping mechanise.
\item \textbf{Weak text-image semantic correspondence:} Images have a significant visual impact on the popularity of a web content. However, the text and images have the weakly semantic correlation to determine the popularity of an event~\cite{li2022grounded}. The second observation is the nonlinear correlations between keywords and images.
\end{itemize}

Motivated by the above two observations, we develop a novel end-to-end deep model termed as Self-Mutual excitation graph Neural network (SMN). The main idea is to map the self-excitation, the mutual excitation among keywords, and the context of images to the popularity score. Concretely, SMN consists of two branches: one branch is the text-based popularity to leverage both self-excitation and mutual excitation among keywords; the another branch is for the image-based popularity prediction. The proposed SMN represents words from different events into an unified graph. Self-excitation is achieved by applying the Straight Through Estimator (STE)~\cite{kang2022forget} to select keywords. Mutual excitation models the nonlinear influence of the co-occurrence among words into the popularity score. In the image-based branch, we employ the Contrastive Language-Image Pre-training(CLIP)~\cite{radford2021learning} to consider the correspondence with the text semantics. The scores from the two branches are added to predict the popularity scores of events. In a nutshell, our contributions are as follows:
\begin{itemize}
    \item To our best knowledge, we firstly propose the self-excitation and the mutual-excitation to model the popularity scores among keywords for the popularity scores. We find that mutual excitation is very critical to popularity score prediction. This motivates us to decompose the popularity score of an event as an additive model.
    \item We propose an interpretable deep factorization method to predict the popularity score for a event on the web. Specifically, our model extracts sparse yet important keywords, which further interpret why a event has a higher popularity score.
    \item We conduct comprehensive experiments on two real-world datasets, demonstrating that SMN outperforms State-Of-The-Art (SOTA) methods, validating the benefits of its components, and providing qualitative analysis for case studies. 
\end{itemize}

\section{Related Work}

\subsection{Popularity Prediction}
A substantial amount of research has focused on understand online video~\cite{pinto2013using}, offline activity~\cite{wang2018factorization}, social text~\cite{cui2011should}, academic paper~\cite{xiao2016modeling}, and multi-modal social image~\cite{zhang2018user}. These research is roughly categorized into two groups: 1) the cold-start method just depends on the content of an event itself~\cite{cui2011should}~\cite{dimitrov2017makes}~\cite{zhang2018user}; and 2) the warm-start method requires historical information provided by early popularity measures to predict the aging effect and triggering of popularity~\cite{pinto2013using,xiao2016modeling,cao2017deephawkes,sriram2010short}. 
For example, by taking advantage of the point process for continuous time modeling, Zhao et al. \cite{zhao2015seismic} introduced the self-exciting point process to predict the final number of reshapes of a post. Liu et al. \cite{yang2016stacked} proposed a feature-based point process to answer natural language questions from images. In this paper, we focus on predicting the future popularity of an event. Our method belongs to the cold-start method.

%https://dl.acm.org/doi/pdf/10.1145/3178876.3186026
In recent years, visual modality has received increasing attention in popularity prediction research~\cite{chen2016micro,wu2017sequential}.
For instance, Chen et al. \cite{chen2016micro} introduced a transductive multi-modal learning model to identify an optimal latent space from these different modalities. However, it is difficult to extended to online prediction. Wu et al. \cite{wu2017sequential} proposed to integrates both temporal context and temporal attention. However, these models are lack interpretability.
In this paper, we explain why an event on the web is more interesting than the others from both the unstructured textual and the visual modalities. 

\subsection{Deep Multi-modal Learning}
There has been increasing attention on using multi-modality to jointly model the popularity prediction. Multi-modal learning concentrates on learning from multiple sources with different modalities \cite{zhang2017uncovering}.
Deep multi-modal learning models involve three types of settings: 1) multi-modal fusion~\cite{nojavanasghari2016deep},
2) cross modality learning~\cite{van2018learning,tan2019lxmert}, and 3) shared representation learning~\cite{kang2012deep}.
Our problem belongs to the multi-modal fusion setting. 

Multi-modal fusion has achieved great success in various tasks, \textit{e.g.}, visual question answering (VQA) \cite{antol2015vqa}, image captioning \cite{karpathy2015deep} and popularity prediction \cite{keneshloo2016predicting,yang2016stacked}, developing from early simple multi-modal fusion to later more complex deep methods. However, to our knowledge, none of multi-modal deep learning methods has been proposed to the popularity prediction task in an interpretable approach, which motivates us to take a step towards this aim.

\subsection{Interpretability of Multi-modal Data}

Because of the over-parameterization and the extreme non-linearity of deep learning models, it is still difficult to interpret their behaviors and outcomes.
An interpretation algorithm is trustworthy if it properly reveals the underlying rationale of a model making decisions \cite{lipton2018mythos}.
The model interpretability refers to the intrinsic properties of a deep model measuring in which degree the inference result of the deep model is predictable or understandable to human beings~\cite{li2022interpretable}.
Technically, interpretability is coarsely summarized as two aspects: 1) hard sparsity or soft sparsity (or attention mechanism)~\cite{mnih2014recurrent}, and 2) uncorrelated prototypes~\cite{ragno2022prototype}~\cite{pang-hu-protptype-t-cybernetics-19}. 
For instance, to select important regions from images, \cite{mnih2014recurrent} focuses on some specific words relevant to machine. In this paper, we propose to generate the weight-related sparsity vis STE for neural network.

\section{METHODOLOGY}\label{sec:methodology}

\subsection{Problem Definition}

Given a posted event in a webpage $\mathcal{X}$, the textual content $\mathcal{T}$ and the images $\mathcal{I}$, \textit{i.e.}, $\mathcal{X}=\{\mathcal{T}, \mathcal{I}\}$ can be extracted from $\mathcal{X}$. Let $\mathcal{T}_i = \{w_1, \ldots, w_{N_i} \}$ denotes a set of words $w_j$ ($1\leq j \leq N_i$) from the textual content of the $i$-th webpage; while the images $\mathcal{I}= \{I_1,\ldots, I_{M_i}\}$ , where $M_i$ is the count of images. Each webpage has a popularity score, \textit{e.g.}, $s$ for $\mathcal{X}$.

Based on the above notations, we formally state the problem as follows:

\begin{mypro}[Popularity prediction of an event]
Given an event $\mathcal{X}$ in a webpage described by $(\mathcal{T,I})$ and its popularity score $s$, our aim is to learn an interpretable function $f(\mathcal{T,I})$ that predict the popularity score, \textit{i.e.}, $f(\mathcal{T,I}) \rightarrow s$.
\end{mypro}

In what follows, for simplicity, we will omit the superscript $_i$ of related notations later. Besides, we use the terms, \textit{i.e.}, embedding and representation, interchangeably.

\subsection{Model}

\begin{figure}[t!]
    \centering
    \includegraphics[scale=0.27]{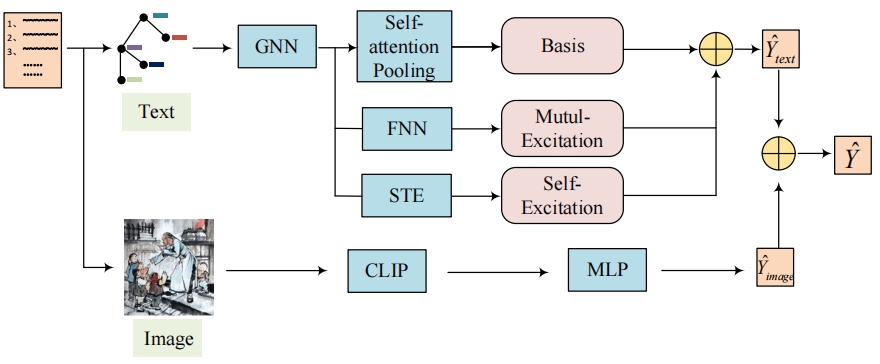}
    \caption{SMN Model System Framework}
    \label{fig3}
\end{figure}

\subsubsection{Represent words into a Graph}

Given a set of events $\{\mathcal{X}_1,\ldots,\mathcal{X}_K\}$, we extract all these tokened words $ w_i \in \mathcal{T}_m\cup \mathcal{T}_n $ ($1\leq ,m\neq n, \leq N$) from all webpages. We extracted Point-wise Mutual Information (PMI)~\cite{church1990word} between two words $w_{i}$ and $w_{j}$ as follows:
\begin{equation}\label{eqt:pmi}
\text{PMI}(i,j)=log \left( \frac{d(i,j)}{\underset{i=1}{\overset{N}{\mathop{\Sigma }}}\,\underset{j=1}{\overset{N}{\mathop{\Sigma }}}\,d(i,j)} \frac{\underset{i=1}{\overset{N}{\mathop{\Sigma }}}\,d(i)\underset{j=1}{\overset{N}{\mathop{\Sigma }}}\,d(j)}{d(i)d(j)} \right)
%\frac{d(i,j)/\underset{i=1}{\overset{N}{\mathop{\Sigma }}}\,\underset{j=1}{\overset{N}{\mathop{\Sigma }}}\,d(i,j)}{(d(i)/\underset{i=1}{\overset{N}{\mathop{\Sigma }}}\,d(i))\cdot (d(j)/\underset{j=1}{\overset{N}{\mathop{\Sigma }}}\,d(j))},
\end{equation}
where $d(i,j)\in {{\mathbb{R}}^{1}}$ is the counts of both the $i$-th and the $j$-th words in all events, $d(i)\in {{\mathbb{R}}^{1}}$ and $d(j)\in {{\mathbb{R}}^{ 1}}$ are the frequencies of the $i$-th and the $j$-th words in all events, respectively, and $N$ is the number of words. A higher PMI score in Eq.~\eqref{eqt:pmi} means that the semantic relevance between two words in the corpus is high. 
The Eq.~\eqref{eqt:pmi} can generate a PMI sparse matrix, which we denote as $\mathbf{A} \in {{\mathbb{R}}^{N\times N}}$.
Therefore, the matrix $\text{PMI}(i,j) \in {{\mathbb{R}}^{N\times N}}$ is a sparse matrix which can be efficiently stored, computed, and updated. 

Let $G=(\mathbf{V},\mathbf{E},\mathbf{A})$ be a weighted graph, where vertex set $\mathbf{V}$ is the union of all words $\mathbf{V}=\{w_i\}, i\in [1,N]$, the affinity matrix $\mathbf{A}$ is equal to the $\text{PMI}(i,j)$ matrix, and the edge set $\mathbf{E}(i,j)$ indicates whether the words $w_i$ and $w_j$ simultaneously occur in a webpage or not. The word embedding vector $\mathbf{h}_i$ for the word $w_i$ is obtained by the Word2vec~\cite{mikolov2013efficient} method.

The motivations behind represent the words into a graph is two folds: 1) Human can quickly interpret the semantic meanings of a webpage by either extracting solo words or multiple words within an event; 2) An unified graph is helpful to capture inter-topic correlation. We further summarize some notations used in this paper in Table~\ref{tab:notation}.

\begin{table}[t!]
\centering
  \caption{Some important notations used in this paper.}
\scalebox{0.85}{
\begin{tabular}{cc}
        \hline

        Notation& Explanation\\
        \hline
        $\mathbf{A}$& adjacency matrix of the graph \\
        $\mathbf{V}$& the set of nodes \\
        $\mathbf{E}$& the set of edges \\
        $\mathbf{S}$& self-attention score matrix for all nodes \\
        $\mathbf{S}_{mask}$& self-attention mask matrix consisted by 0 and 1 \\
        $\mathbf{\tilde{H}}$& nodes feature matrix from self-attention pooling\\
        $\mathbf{\hat{H}}$& nodes feature of excited state \\
        $\mathbf{X}_{image}$& image feature matrix extracted by CLIP \\
        $\mathbf{W}_{mask}$& the sparsity mask matrix consisted by 0 and 1 \\
        \hline
    \end{tabular}}
  \label{tab:notation}
\end{table}

\subsection{Problem Formulation of SMN}

\subsubsection{Selection of Keywords} Given a graph $G$ built from all events webpages and the initial words embedding $\mathbf{h}_i$, the embeddings $\mathbf{h}_i$ are concatenated into a matrix $\mathbf{H}=[{\mathbf{h}_{1}},...,\mathbf{{h}}_{i},...,{\mathbf{h}_{N}}] \in {{\mathbb{R}}^{N\times F}}$, where $F$ is the dimension of the embedded vector. In this paper, we taken GCN~\cite{kipf2016semi} as backbone where the embedded word representation is computed as follows:
\begin{equation}
\centering
{\mathbf{H}^{(l+1)}}=relu \big({{\mathbf{\tilde{D}}}^{-\frac{1}{2}}}\mathbf{\tilde{A}}{{\mathbf{\tilde{D}}}^{-\frac{1}{2}}}{\mathbf{H}^{(l)}}{\mathbf{W}^{(l)}}\big),
\end{equation}
where $\mathbf{\tilde{A}}=\mathbf{A}+\mathbf{I}$ is the adjacency matrix with added self-connection matrix $\mathbf{I}\in\mathbb{R}^{N\times N}$, the degree matrix $\mathbf{\tilde{D}}={{\Sigma }_{j}}{\mathbf{\tilde{A}}_{ij}}$. $\mathbf{{H}}^{(l)} \in {{\mathbb{R}}^{N\times F}}$ is the feature of the ${l}$-th layer for all words, and $\mathbf{{W}}^{(l)}\in {{\mathbb{R}}^{N\times {F}}}$ is the weight matrix of the $l$-th layer. 

To obtain the event-level features for each event, we add the self-attention pooling operation~\cite{lee2019self}, which softly distinguishes whether a node should be deleted or not. Specifically, self-attention scores are computed by graph convolution as follows:
\begin{equation}
\mathbf{S}=\sigma ({\mathbf{\widetilde{D}}^{-\frac{1}{2}}}\mathbf{\widetilde{A}}{\mathbf{\widetilde{D}}^{-\frac{1}{2}}}{\mathbf{{H}}^{(l+1)}}\mathbf{\Theta} ),
\end{equation}
where $\mathbf{\Theta} \in {\mathbb{R}}^{F\times F_{c}}$ is the parameter of the pooling layer, and $F_{c}$ is the dimension of the output feature. $\mathbf{S} \in {{\mathbb{R}}^{N\times F}}$ is the score matrix of all words. Based on soft attention mask, the node selection process is defined as follows:
\begin{align}
\centering
idx &= \text{top-rank}(\mathbf{S},kN),\\
{\mathbf{S}_{mask}} &= {\mathbf{S}_{idx}},
\end{align}
where the pooling rate ratio $k\in (0,1]$ is a hyperparameter that determines the number of nodes retained, function $\text{top-rank}(\cdot)$ returns the index of the top $kN$ nodes, $idx$ is the returned index, and $\mathbf{S}_{mask} \in {{\mathbb{R}}^{N\times F}}$ is the feature attention mask. Therefore, the node matrix can be expressed as follows:
\begin{equation}\label{eqt:indexed_word_embedding}
\centering
\mathbf{\tilde{H}}^{(l+1)}={\mathbf{H}^{(l+1)}}\odot {\mathbf{S}_{mask}},
\end{equation}
where $\odot$ is the Hadamard product, which represents the multiplication of elements at corresponding positions.

%\subsubsection{Image Representation}

% CLIP Contextual Image Representation for Popularity

\subsubsection{Mutual Excitation, Self Excitation and Base Excitation of Keywords}

Mutual excitation mechanism reflects whether multiple keywords would increase a larger interestingness value than itself to indicate the popularity of the content. For example, the event ``The resale prices of KFC blind box products have surged 8 times'' has a high popularity score due to the simultaneous occurrence of ``KFC'' and ``blind box''.

We firstly project the feature of words in Eq.~\eqref{eqt:indexed_word_embedding} into the embedded excitation space as follows:
\begin{equation}
 {\mathbf{\hat{H}}}^{(l+1)}= relu\text{(}{\mathbf{W}_{2}}(relu({\mathbf{W}_{1}}{\mathbf{H}^{(l+1)}})) ,
\end{equation}
where $\mathbf{{H}}^{(l+1)} \in {{\mathbb{R}}^{N\times F}}$ is the output of the last GCN layer. 

We model the mutual excitation process through an exponential linear layer as follows:
\begin{equation}
\hat{y}_{m}^{i}
 ={{\eta }_{i}}\sum\limits_{j}\sum\limits_{k}{\exp (-{{\gamma }_{i}}\parallel \mathbf{\hat{h}}_{j}-\mathbf{\hat{h}}_{k}\parallel _{2}^{2})},
\end{equation}
where $\hat{y}_{m}^{i}\in {{\mathbb{R}}^{1}}$ is the mutual-excitation popularity of the $i$-th event. ${{\eta }_{i}}$ captures the excitation and inhibition effects between words, and ${{\gamma }_{i}}$ is the coefficient of the $i$-th event word excitation term, which be calculated as follows:
\begin{equation}
\centering
{{\eta }_{i}}=gelu({\mathbf{\hat{h}}_{i}}{\mathbf{W}_{\eta }}), 
{{\gamma }_{i}}=gelu({\mathbf{\hat{h}}_{i}}{\mathbf{W}_{\gamma }}).   
\end{equation}

%where ${\mathbf{\hat{h}}_{i}}$ is the $i$-th event after self-attention pooling feature, same as ${\mathbf{\hat{h}}_{k}}$. 
Specifically, $\parallel \mathbf{\hat{h}}_{j}^{{}}-\mathbf{\hat{h}}_{k}^{{}}\parallel _{2}^{2}$ is calculated as follows:
\begin{equation}
\centering
\begin{aligned}
\parallel\mathbf{\hat{h}}_{j}^{{}}-\mathbf{\hat{h}}_{k}^{{}}\parallel_{2}^{2}
 &=(\mathbf{\hat{h}}_{j}^{{}}-\mathbf{\hat{h}}_{k}^{{}}){{(\mathbf{\hat{h}}_{j}^{{}}-\mathbf{\hat{h}}_{k}^{{}})}^{T}} \\
 &=\parallel \mathbf{\hat{h}}_{j}^{{}}\parallel _{2}^{2}+\parallel \mathbf{\hat{h}}_{k}^{{}}\parallel _{2}^{2}-2\underbrace{\mathbf{\hat{h}}_{j}^{{}}\mathbf{\hat{h}}{{_{k}^{{}}}^{T}}}_{z_{jk}}
\end{aligned}
\end{equation}
where ${z_{jk}}$ describes the mutual excitation between the two words ${w}_{j}$ and ${w}_{k}$.

Self-excitation models the popularity of each key words. Specifically, the self-excitation ${\beta }_{ij}$ of the $j$-th keyword of the $i$-th webpage is as follows:
\begin{equation}\label{eqt:self_excitation}
\centering
{{\beta }_{ij}}\text{  }=\text{ }gelu({\mathbf{h}_{ij}}{\mathbf{W}_{\beta }}),
\end{equation}
where ${\mathbf{h}_{ij}}\in {{\mathbb{R}}^{F\times {1}}}$ is the feature vector for $j$-th word of textual content in the $i$-th webpage, ${\mathbf{W}_{\beta }}\in {{\mathbb{R}}^{F\times {1}}}$ is weight. The self-excited popularity in $i$-th webpage is calculated as follows:
\begin{equation}
\hat{y}_{s}^{i}\text{  }=\sum\limits_{j}{{\beta }_{ij}}.
\end{equation}

The base popularity score is assumed to predict the popularity score from the topics of a event. Therefore, the pooled embedding vectors are used to predict popularity as follows:
\begin{equation}
\hat{y}_{b}^{i}=gelu({\mathbf{\tilde{h}}_{i}}{\mathbf{W}_{\mu }}),
\end{equation}
where $\mathbf{W}_{\mu }\in {{\mathbb{R}}^{F\times {1}}}$ is a linear layer parameter, ${\mathbf{\tilde{h}}}_{i}$ is the feature after self-attention pooling of Eq.~\eqref{eqt:indexed_word_embedding}.

\subsubsection{Contextual Image Representation}

The CLIP contains text encoder function $E_{text}$ and image encoder function $E_{image}$. Given an image ${{I}_{i}}$ from the $i$-th webpage, the feature extraction is as follows:
\begin{equation}
\mathbf{x}_{image}^{i}={E_{image}}({{I}_{i}}),
\end{equation}
where $\mathbf{x}_{image}^{i} \in {{\mathbb{R}}^{1\times {F}_{c}}}$ is the extracted image feature vector of the $i$-th webpage, which is further fed into the MLP to obtain the popularity of the $i$-th webpage $\hat{y}_{image}^{i} \in {{\mathbb{R}}^{1}}$, as follows:
\begin{equation}
\hat{y}_{image}^{i}=\text{MLP}(\mathbf{x}_{image}^{i}),
\end{equation}
where MLP consists of two nonlinear layers in which the $relu$ activation function is used to learn mapping relationships between image and text features based on multiple hidden layers.

\subsubsection{Additive Popularity Prediction Mechanism}

The popularity of the $i$-th webpage is defined as follows:
\begin{equation}
\hat{y}^{i}=\hat{y}_{b}^{i}+\hat{y}_{s}^{i}+\hat{y}_{m}^{i} + \hat{y}_{image}^{i},
\end{equation}
where $\hat{y}_{b}^{i}$, $\hat{y}_{s}^{i}$, $\hat{y}_{m}^{i}$, and $\hat{y}_{image}^{i}$ are the basis, self-excitation, mutual-excitation, image popularity, respectively.

\subsection{Interpretability via Sparsity}

The sparsity of words would increase the interpretability of a webpage~\cite{pang2018increasing}. We filter the unimportant keywords by sparing the parameters. Concretely, we obtain the mask matrix ${\mathbf{W}_{mask}}$ by a percentage threshold $\delta$ to set the top $\delta$$\%$ of values in ${\mathbf{W}_{\beta}}$ to 1 and the others to 0 as follows:
\begin{equation}
\begin{aligned}
{\mathbf{W}_{mask}}
& =f({\mathbf{W}_{\beta}}) \\
& =\left\{ \begin{aligned}
  & 1,\text{      }{{w}_{\beta}}\geq {{v}_{\delta{\%}}}, \\
 & 0,\text{      }  otherwise, \\
\end{aligned} \right.
\end{aligned}
\end{equation}
where ${{w}_{\beta}}$ is the value in ${\mathbf{W}_{\beta}}$, ${{v}_{\delta{\%}}}$ is the $\delta$ minimum value in ${\mathbf{W}_{\beta}}$.

The mask matrix ${\mathbf{W}_{mask}}$ is then multiplied with the matrix ${\mathbf{W}_{\beta}}$ to obtain the new parameters ${\mathbf{W}_{pruned}}$ as follows:
\begin{equation}
{\hat{\beta }_{ij}}=gelu(\mathbf{{h}}_{ij}\big( {\mathbf{W}_{mask}}\odot {\mathbf{W}_{\beta }}\big),
\end{equation}
where $\hat{\beta}_{ij}$ is the filtered self-excitation score, obtained from the weighting parameter with sparsity.

\begin{table*}[t!]
\centering
\caption{Comparison between the proposed Hot Events dataset and Sina
Weibo-1, HEP-PH, Twitter-3, TPIC17.}
\scalebox{0.7}{
\begin{tabular}{cccccc}
    \hline
    Dataset& Data Source& Text & Image & Data Publisher& Popularity\\
    \hline
    Weibo-1 \cite{cao2017deephawkes}& June 1,2016 Sina Weibo& $\surd$ & $\times$ & User&Hotness index\\
    Twitter-3 \cite{zhao2015seismic}& August 8-12,2016 Twitter data& $\surd$ & $\times$ & User& Retweets of twitters\\
    TPIC17 \cite{wu2017sequential}& Flicker platform images of User browsing data& $\surd$ & $\times$ & User&  User Views\\
    HEP-PH \cite{leskovec2005graphs}& 1993-2003 High Energy in Arxiv Physics-related papers& $\surd$ & $\times$& Academic researchers& Number of paper citations\\
    Hot events (our dataset)& October 10-,2021 Baidu daily Hotlist & $\surd$ & $\surd$ & User& Event popularity value\\
    \hline
\end{tabular}}
\label{tab2}
\end{table*}

\subsection{Optimization}

We design the loss function as a combination of the Huber regression loss and sparsity node regularization, as follows:
\begin{equation}
Loss(\hat{y}^{i},y^{i})=Huber(\hat{y}^{i},y^{i})+{{\lambda }_{1}}\parallel {{\beta }_{i}}{{\parallel }_{1}}+{{\lambda }_{2}}\parallel {{z}_{i}}{{\parallel }_{1}},
\end{equation}
where ${{\beta }_{i}}$ and ${{z}_{i}}$ are the self-excitation and mutual-excitation scores of the words, respectively. With ${{\lambda }_{1}}$ and ${{\lambda }_{2}}$ as the regular term coefficients.

The gradient back-propagation process of the parameter $\mathbf{W}_{mask}$ is defined as follows:
\begin{equation}
\frac{\partial Loss(\hat{y}^{i},y^{i})}{\partial {\mathbf{W}_{\beta}}}=\frac{\partial loss}{\partial {\mathbf{W}_{mask}}}\cdot \frac{\partial {\mathbf{W}_{mask}}}{\partial {\mathbf{W}_{\beta}}},
\end{equation}
where requires solving the back-propagation gradient irreducibly problem, the main approach is to make $\frac{\partial {\mathbf{W}_{mask}}}{\partial {\mathbf{W}_{\beta}}} = 1$, that is, the back propagation process uses $\frac{\partial loss}{\partial {\mathbf{W}_{mask}}}$ instead of the original $\frac{\partial loss}{\partial {\mathbf{W}_{\beta}}}$. %The STE quantized binary matrices instead of floating-point parameters for computation during forward propagation. 
In the back propagation process, the derivative of the output is directly used as the derivative of the input to complete the sparse process of the parameters.

\section{Experiments}
%In this section, we evaluate the proposed method on two popularity prediction tasks. Specifically, we want to verify whether the proposed method achieves satisfactory results and whether observing sparse focus words helps to explain our model. We will also show some specific experimental details.

\subsection{Datasets}
We build a Hot Events dataset from Baidu Hotlist~\footnote{https://top.baidu.com/board}. The dataset contains 4,000 events and 4,000 images, with text titles consisting of 6 to 17 words and text content ranging from 50 to 500 words. The data includes hot events details(\emph{i.e.}, event title, content description), hot search index and images. We compared the proposed dataset with four existing datasets (Sina Weibo-1, HEP-PH, Twitter-3 and TPIC17) in Table~\ref{tab2}.

\subsection{Comparison Methods}\label{comparison_methods}

Our experimental goals include two aspects:

\uppercase\expandafter{\romannumeral1}) Understanding the effectiveness of different interaction methods in the proposed method:
\begin{itemize}
\item [1.] \textbf{SMN-GCN.} The backbone of text modality is Graph Convolutional Networks (GCN) \cite{kipf2016semi}, which employs graph convolutional layers to propagate information to improve classification accuracy. 
\item [2.] \textbf{SMN-GAT.} The backbone of text modality is Graph Attention Networks (GAT) \cite{velivckovic2017graph}, which introduces an attention mechanism to graph-based learning, allowing nodes to weigh their neighbors' contributions dynamically.
\end{itemize}

GCN and GAT are two widely used graph neural networks. In the following experiments, we use these two networks as the backbone for the text modality to showcase the performance of the interaction models.

\uppercase\expandafter{\romannumeral2}) We compare the proposed  SMN with the SOTA models as follows: 
\begin{itemize}
   \item  \textbf{Factorization Machines (FM)} \cite{rendle2010factorization}: FM models the correlation between factors and the interactions among variables. 
   \item  \textbf{Field-aware Factorization Machines(FFM)} \cite{juan2016field}: FFM incorporates the concept of fields (or topics) into FM for handling the interaction between features. Therefore, we use both methods as SOTA to compare with SMN, proving the importance of the interaction between features.
   \item  \textbf{DeepFM} \cite{guo2017deepfm}: DeepFM is based on the wide and deep neural network structure, using FM in the wide part to avoid the hand-crafted feature engineering. Compared with FM, DeepFM models both the low-level and the high-level features. 
   \item  \textbf{xDeepFM} \cite{lian2018xdeepfm}: xDeepFM improves the DeepFM. It combines the cross-networks and Factorization Machines, attempting to learn complex feature interactions by DNN. We use DeepFM and xDeepFM as comparison to highlight the role of deep features.
   \item \textbf{Factorisation-machine supported Neural Network (FNN)} \cite{zhang2016deep}: FNN implements both the explicit feature intersection of FM and the implicit higher-order intersection by DNN. 
\end{itemize}

It should be note that the SOTA methods (\textit{i.e.}, FM, FFM, DeepFM and xDeepFM) are originally proposed to deal with the Click-Through Rate (CTR) problem. Although the SOTA methods and our method try to model the one-order and high-order correlations among different factors into the regression tasks. The difference between the popularity prediction for the events on web and the CTR problem is that different domains need the diverse correlations for each task. 

The comparisons between the SOTA methods and our method illustrate that the off-the-shelf methods barely be applied directly to the popularity prediction for the events on web. On the contrary, the proposed base, self and mutual excitation are critical to our task. 

\subsection{Evaluation Metrics}

Four evaluation metrics, Mean-Square Error (MSE), Order Loss (OL), mean Average Precision (mAP), and Discounted Cumulative Gain Normalized (NDCG), are used to assess the proposed method and the comparison methods as follows:
\begin{itemize}
\item \textbf{MSE} measures the difference between the predicted popularity value and the ground truth popularity as follows:
\begin{equation}
MSE=\frac{1}{N}\sum\limits_{i=1}^{N}{\left| {\hat{y}^{i}-y^{i}} \right|}.
\end{equation}
\item \textbf{OL@$\mathbf{K}$} measures the performance of the top $K$ predicted events as follows:
\begin{equation}
OL=\frac{\sum\limits_{k=0}^{K}{|{{R}_{k}}-{{P}_{k}}|}}{T},
\end{equation}
where $T$ is the total number of the hot events, ${R}_{k}$ denotes the ground truth popularity value of the top $k$ event, ${P}_{k}$ denotes the top $k$ predicted popularity one ($0\leq k \leq K$).
Therefore, OL@$K$ focuses on the performance of the top $K$ events.

\item \textbf{mAP} measures the degree of ranking of multiple events as follows:
\begin{equation}
mAP(\mathbf{y} ,l)=\frac{\sum\limits_{k=1}^{K}P@k\cdot {I\{{{l}_{\mathbf{y}_{k}}}=1\}}}{K}, 
\end{equation}
where
\begin{equation}
P@k(\mathbf{y} ,l)=\frac{\sum\nolimits_{t\le k}{{{I}{\{{l_{\mathbf{y}_k }}=1\}}}}}{k},
\end{equation}
in which $\mathbf{y}$ denotes the popularity scores for all events, $I(\cdot)$ is the indicator function, $k$ denotes the first $k$ ranked events. mAP measures how good the popularity scores when a threshold is given.

\item \textbf{NDCG} measures the correctness and the relevance of a ranked list as follows:
\begin{equation}
NDCG(\mathbf{y} ,l)=\frac{1}{{{Z}_{K}}}\sum\limits_{i=1}^{K}{G({{l}_{{{\mathbf{y} }_i}}})\eta (i)},
\end{equation}
where $K$ denotes the number of events with predicted events ranked in the top $K$, $\eta (i)=\frac{1}{lo{{g}_{2}}(i+1)}$ is the discount factor that comes with a lower sorting position, $G(\cdot)$ is used for scoring popularity values, ${Z}_{K}$ is the DCG when the model ranking result is optimal. 
\end{itemize}

\subsection{Implementation Details}
%In order to unify the performance of the metric model under each evaluation index, 
We normalize all the popularity scores into the range [0, 1] for the numerical stability. 
%We randomly select 1,000 hot events, using 800 of them as the training set, which contains approximately 3,000 words. The remaining 200 are used as test data, and contains about 700 words.
The encoder of CLIP is selected from the pre-training model ``ViT-B/32''. The loss function regular term coefficient ${{\lambda }_{1}}$=${{\lambda }_{2}}$=0.001. The learning rate is set to 0.01. The Stochastic Gradient Descent (SGD) optimizer is used for iterative optimization to complete all the experiments. Meanwhile, in order to make the network model converge to the optimal solution, we also add the cosine annealing warm restarts strategy\cite{loshchilov2016stochastic}. We select six different mAP thresholds $m$=$\{6,7,8,9,10,15\}$ and set $k$=10 in the NDCG metric.

%The effectiveness of GNNs directly impact the performance of text prediction. Therefore, we chose GCN and Graph Attention Network (GAT) \cite{velivckovic2017graph} as the backbone networks for the SMN text branch, referred to as SMN-GCN and SMN-GAT, respectively.

\subsection{Hyper-parametric Analysis}

\begin{figure}[t]
    \centering
    \includegraphics[scale=0.3]{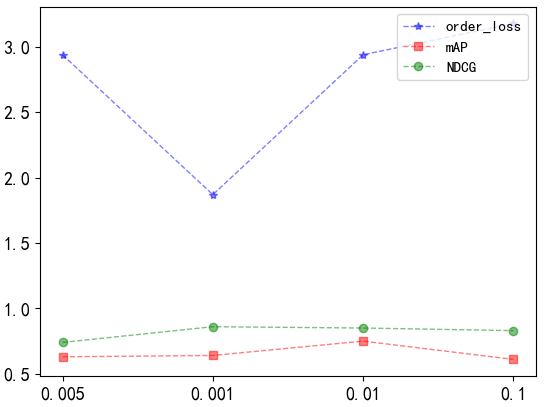}
    \caption{ Results of SMN-GAT on three metrics (OL@10, mAP, NDCG) with four different sets of coefficient settings on Hot Events dataset.}
    \label{fig4}
\end{figure}

\begin{figure}[t]
    \centering
    \includegraphics[scale=0.3]{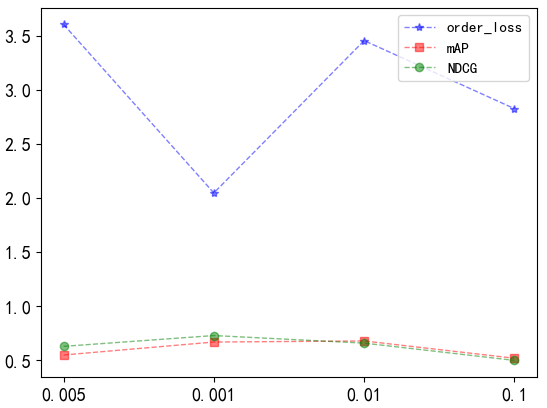}
    \caption{Results of SMN-GAT on three metrics (OL@10, mAP, NDCG) with four different sets of coefficient settings on HEP-PH dataset.}
    \label{fig5}
\end{figure}

In order to explore the effect of the regular term coefficients ${{\lambda }_{1}}$ and ${{\lambda }_{2}}$, we first make that ${{\lambda }_{1}}$ is equal to ${{\lambda }_{2}}$, and search for the optimal setting in the range $\{0.1, 0.01, 0.001, 0.005\} $. The results are shown in Fig.~\ref{fig4} and Fig.~\ref{fig5}.
 

Figs.~\ref{fig4} and~\ref{fig5} discover that the OL exhibits significant fluctuations with the best performance when the regular terms are 0.001. In contrast, the mAP and NDCG metrics display a stable performances with respect to the change of the regular terms. The results indicate that our model has good generalization ability across different tasks; besides, the hyper-parameters are relative stable for different tasks. Therefore, we set both ${{\lambda }_{1}}$ and ${{\lambda }_{2}}$ is equal to 0.001, respectively in our experiments.

\begin{table}[h!]
  \centering
  \caption{Performance comparison of image and text modality in SMN-GCN and SMN-GAT on Hot events dataset. $\checkmark$ indicates which modality is used. The best accuracy is in bold, and the second best is underlined.}
  %\resizebox{\linewidth}{!}{
\scalebox{0.6}{
    \begin{tabular}{c|cc||c|ccc|cc}
        \hline
        Model& Image & Text & MSE $\downarrow$& OL@10 $\downarrow$& OL@20$\downarrow$& OL@30$\downarrow$& mAP $\uparrow$& NDCG $\uparrow$\\
        \hline\hline
        $\text{SMN}$&$\checkmark$&& \underline{0.09}& 3.77& 7.38& 10.89& 0.64& 0.68\\\hline
        $\text{SMN-GCN}$&&$\checkmark$&0.13&2.74&7.37&10.70& 0.68& 0.71\\
        $\text{SMN-GAT}$&&$\checkmark$&\textbf{0.08}&2.72&\underline{6.14}&10.75& 0.65& 0.71\\ \hline
        $\text{SMN-GCN}$&$\checkmark$&$\checkmark$& 0.10& \underline{2.55}& 6.60& \underline{9.79}& \underline{0.69}& \underline{0.83}\\
        $\text{SMN-GAT}$&$\checkmark$&$\checkmark$& \underline{0.09}& \textbf{1.87}& \textbf{5.13}& \textbf{8.57}& \textbf{0.75}& \textbf{0.85}\\
        \hline
    \end{tabular}}
  \label{influ_modality}
\end{table}

\subsection{Ablation Study}\label{sec:sub:ablation}

\textbf{Influence of Different modalities:} We separately remove the image and text modalities on our method to verify the impact of different modalities.
We first compare the effectiveness of the image modality and the text modality. Table \ref{influ_modality} shows that the effectiveness of the text modality is slightly outperforms the image modality. For instance, SMN-GAT with only text modality improves by 0.01, 1.05, 0.01, and 0.03 in MSE, OL@10, mAP, and NDCG compared to SMN with only the image modality. This indicates that the image modality contains the useful cue to predict the popularity scores of events from web.   

Another observation is that the performance of our method is significantly improved when both image and text modalities are used. For instance, SMN-GAT with the image and the text modalities improves by 1.9, 0.1, and 0.17 in OL@10, mAP, and NDCG compared to SMN with the image modality. This indicates that the fusion of the multiple modalities is better than a single modality. 

\begin{table}[t!]
  \centering
  \caption{Performance comparison of different interaction models (Base, Self, Mutual) on Hot Events dataset. $\checkmark$ indicates which excitation model is used.}
  %\resizebox{\linewidth}{!}{
\scalebox{0.55}{
    \begin{tabular}{c|ccc||c|ccc|cc}
        \hline
        Model&Base &Self &Mutual & MSE $\downarrow$& OL@10 $\downarrow$& OL@20$\downarrow$& OL@30$\downarrow$& mAP $\uparrow$& NDCG $\uparrow$\\
        \hline\hline
        \multirow{4}{*}{SMN-GCN}&$\checkmark$&$\checkmark$&$\checkmark$& 0.10& \underline{2.55}& \underline{6.60}& \underline{9.79}& \underline{0.69}& \underline{0.83}\\
        &$\checkmark$&&$\checkmark$&\underline{0.09}&2.92&8.28&13.35& 0.61& 0.69\\
        &$\checkmark$&$\checkmark$&&0.25&3.09&7.54&11.08& 0.61& 0.66\\
        &$\checkmark$&&&\textbf{0.08}&3.88&8.76&12.17& 0.56& 0.62\\
        \hline

        \multirow{4}{*}{SMN-GAT}&$\checkmark$&$\checkmark$&$\checkmark$& \underline{0.09}& \textbf{1.87}& \textbf{5.13}& \textbf{8.57}& \textbf{0.75}& \textbf{0.85}\\
        &$\checkmark$&&$\checkmark$&\underline{0.09}&3.38&7.53&10.74& 0.62& 0.69\\
        &$\checkmark$&$\checkmark$&&\underline{0.09}&2.59&7.69&11.79& 0.67& 0.73\\
        &$\checkmark$&&&0.10&4.85&11.24&14.72& 0.55& 0.61\\
                \hline
    \end{tabular}}
  \label{six_variants_hotevents}
\end{table}

\begin{table}[t!]
\centering
\caption{Performance comparison of different interaction models (Base, Self, Mutual) on HEP-PH dataset. $\checkmark$ indicates which excitation model is used.}
%\resizebox{\linewidth}{!}{
\scalebox{0.55}{
\begin{tabular}{c|ccc||c|ccc|cc}
    \hline
        Model&Base &Self &Mutual & MSE $\downarrow$& OL@10 $\downarrow$& OL@20$\downarrow$& OL@30$\downarrow$& mAP $\uparrow$& NDCG $\uparrow$\\
        \hline\hline
    \multirow{4}{*}{SMN-GCN}&$\checkmark$&$\checkmark$&$\checkmark$& 0.94& 2.47& 7.90& 12.57& \underline{0.71}& \textbf{0.73}\\
    &$\checkmark$&&$\checkmark$& \underline{0.23}& 2.68& 7.39& 12.34& 0.58& 0.56 \\
    &$\checkmark$&$\checkmark$&& 0.24& 2.59& 7.37& \textbf{11.81}& 0.61& 0.68 \\
    &$\checkmark$&&& \textbf{0.01}& \textbf{1.80}& 7.16& \underline{12.03}& \textbf{0.74}& 0.69 \\
    \hline
    \multirow{4}{*}{SMN-GAT}&$\checkmark$&$\checkmark$&$\checkmark$& \underline{0.23}&2.05& \textbf{6.41}& 12.11& 0.67& \textbf{0.73}\\
    &$\checkmark$&&$\checkmark$& \underline{0.23}& \underline{1.98}& \underline{6.89}& 12.14& 0.63& \underline{0.72} \\ 
    &$\checkmark$&$\checkmark$&& 0.24& 2.39& 7.62& 12.61& 0.66& 0.67 \\
    &$\checkmark$&&& \textbf{0.01}& \textbf{1.80}& 7.16& \underline{12.03}& 0.70& 0.65 \\
    \hline
\end{tabular}}
  \label{six_variants_hepph}
\end{table}

\textbf{Influence of the different excitation modules:} To verify the impact of the different excitation modules in our model, we separately remove the self excitation, the mutual excitation, the self excitation and mutual excitation from the text branch.
Table~\ref{six_variants_hotevents} and Table~\ref{six_variants_hepph} details the performance of our method across different datasets, uncovering the 3 observations as follows: 
\begin{itemize}
    \item \textit{For the Hot Events dataset, all excitation modules together would bring the best performance for our method over all evaluation metrics.} For instance, SMN-GAT respectively improves by 0.01, 2.98, 0.2, and 0.24 in MSE, OL@10, mAP, and NDCG, compared to SMN-GAT with only the base excitation. It indicates that these three modules work together to extract the global (\textit{i.e.}, base excitation) feature, as well as both the one-order (\textit{i.e.}, self excitation) and the high-order (\textit{i.e.}, mutual excitation) features to build the embedding representation of key words.
    \item \textit{For the HEP-PH dataset, the base excitation module obtains the best performances over most of the evaluation metrics.} For instance, SMN-GCN with only the base excitation improves by 0.93, 0.67, 0.54, and 0.03 in MSE, OL@10, OL@30, and mAP compared to SMN-GAT, respectively. This might attribute to the different text contents of the different datasets. Specially, the HEP-PH dataset consists of the academic papers where the popular works are mainly influenced by the interesting yet important research topics. Consequently, the base excitation module has a better performance than that of both the mutual and self excitation. 
     \item \textit{Interestingly, SMN-GAT obtains the better performances than SMN-GCN on the Hot Events dataset. Conversely, SMN-GCN obtains the better performances than SMN-GAT on the HEP-PH dataset}. We notice that the events on web are very diverse, as illustrated in Fig.~\ref{fig:examplebaidu} and Table~\ref{tab1:keywords}. Therefore, the key words from the different popular events on web barely have an intersection. For example, the key words in the event about \textit{pyramid scheme} (\#3) in Table~\ref{tab1:keywords} is totally different to the \textit{privacy breaching} (\#2) in Table~\ref{tab1:keywords}. Besides, the attention mechanism in GAT is more conducive to diverse vocabulary modeling than GCN. 
\end{itemize}

\begin{table}[t!]
\centering
  \caption{The comparisons between the STOA methods and our method on the Hot events dataset. The best accuracy is in bold, and the second best is underlined.}
\resizebox{\linewidth}{!}{
%\scalebox{0.75}{
\begin{tabular}{c|c|ccc|cc}
        \hline
        Model& MSE $\downarrow$& OL@10 $\downarrow$& OL@20$\downarrow$& OL@30$\downarrow$& mAP $\uparrow$& NDCG $\uparrow$\\
        \hline\hline
        FM& 1.41& 5.18& 9.71& 12.67& 0.49& 0.63\\\hline
        FFM& 1.35& 4.69& 9.88& 11.34& 0.52& 0.66\\
        FNN& 14.93& 4.65& 9.20& 14.66& 0.52& 0.67\\
        DeepFM& 1.03& 3.98& 7.96& 11.04& 0.59& 0.70\\
        xDeepFM& 1.02& 3.41& 7.83& 12.15& 0.63& 0.74\\ \hline
        $\text{SMN-GCN}$& \underline{0.10}& \underline{2.55}& \underline{6.60}& \underline{9.79}& \underline{0.69}& \underline{0.83}\\
        $\text{SMN-GAT}$& \textbf{0.09}& \textbf{1.87}& \textbf{5.13}& \textbf{8.57}& \textbf{0.75}& \textbf{0.85}\\
        \hline
    \end{tabular}}
  \label{result_hot}
\end{table}

\subsection{Comparison with SOTA methods}

\subsubsection{Results on Hot Events}

We compare our method with several
SOTA methods with different evaluation metrics on the Hot Events dataset. Table~\ref{result_hot} shows that our method outperforms all other SOTA methods in all the  evaluation metrics. Note that we construct two variants
of our method (\textit{i.e.}, SMN-GCN and SMN-GAT) with two kinds of GNNs. For instance, compared to the shallow model(\textit{i.e.}, FM), the proposed SMN-GAT achieves improvements of 1.32, 3.31, 0.26, and 0.22 in MSE, OL$@10$, mAP, and NDCG, respectively. When we compare with the deep learning based methods (\textit{i.e.}, FFM, FNN, DeepFM, and xDeepFM), the SMN-GAT still obtains the significant improvements of 0.93, 1.54, 0.12, and 0.11 in MSE, OL$@10$, mAP, and NDCG compared with xDeepFM, respectively. The results generally highlight that the deep learning based method is better than the shallow one to model the non-linear interaction among key words for the popularity prediction tasks.

\begin{CJK*}{UTF8}{gbsn}
\begin{table*}[t!]
\scriptsize 
\centering
\caption{Four examples are used to show the relationship between keywords and their sores with respect to the popularity index. The score for each word is calculated from the self-excitation and the popularity is normalized by the popularity search index in Fig. \ref{fig:examplebaidu}.}\label{tab1:keywords}
\begin{tabular}{c|cccccccc|c}
\hline
No. & \multicolumn{8}{c}{Key word and its self excitation score}       & Popularity \\ \hline\hline
1   & \begin{tabular}[c]{@{}c@{}}二手\\ (second-hand)\\ 0.66\end{tabular}         & \begin{tabular}[c]{@{}c@{}}肯德基\\ (KFC)\\ 0.56\end{tabular}        & \begin{tabular}[c]{@{}c@{}}暴涨\\ (boom)\\ 0.47\end{tabular}          & \begin{tabular}[c]{@{}c@{}}价格\\ (price)\\ 0.44\end{tabular}   & \begin{tabular}[c]{@{}c@{}}盲盒\\ (mystery boxes)\\ 0.29\end{tabular}   & \begin{tabular}[c]{@{}c@{}}平台\\ (platform)\\ 0.29\end{tabular} & \begin{tabular}[c]{@{}c@{}}倍\\ (times)\\ 0.00\end{tabular}       & \begin{tabular}[|c]{@{}c@{}}元\\ (yuan)\\ 0.00\end{tabular}   & 0.87       \\\hline
2   & \begin{tabular}[c]{@{}c@{}}王冰冰\\ (Wang Bingbing)\\ 0.77\end{tabular}      & \begin{tabular}[c]{@{}c@{}}人肉\\ (doxxing)\\ 0.77\end{tabular} & \begin{tabular}[c]{@{}c@{}}隐私\\ (privacy)\\ 0.68\end{tabular}       & \begin{tabular}[c]{@{}c@{}}新娘\\ (bride)\\ 0.67\end{tabular}   & \begin{tabular}[c]{@{}c@{}}照片\\ (photo)\\ 0.40\end{tabular}      & \begin{tabular}[c]{@{}c@{}}露出\\ (privacy breaches)\\ 0.33\end{tabular}  & \begin{tabular}[c]{@{}c@{}}曝光\\ (exposure)\\ 0.09\end{tabular}   & \begin{tabular}[c]{@{}c@{}}媒体\\ (media)\\ 0.00\end{tabular} & 0.78       \\\hline
3   & \begin{tabular}[c]{@{}c@{}}法规\\ (regulation)\\ 0.61\end{tabular}          & \begin{tabular}[c]{@{}c@{}}传销\\ (pyramid scheme)\\ 0.45\end{tabular} & \begin{tabular}[c]{@{}c@{}}禁止\\ (prohibit)\\ 0.41\end{tabular}      & \begin{tabular}[c]{@{}c@{}}夫妇\\ (couples)\\ 0.29\end{tabular} & \begin{tabular}[c]{@{}c@{}}张庭\\ (Zhang Ting)\\ 0.20\end{tabular} & \begin{tabular}[c]{@{}c@{}}案件\\ (case)\\ 0.15\end{tabular}    & \begin{tabular}[c]{@{}c@{}}规避\\ (evade)\\ 0.09\end{tabular}      & \begin{tabular}[c]{@{}c@{}}河北\\ (Hebei)\\ 0.00\end{tabular} & 0.54       \\\hline
4   & \begin{tabular}[c]{@{}c@{}}金球奖\\ (Golden Globe Award)\\ 0.51\end{tabular} & \begin{tabular}[c]{@{}c@{}}获奖\\ (award)\\ 0.46\end{tabular}       & \begin{tabular}[c]{@{}c@{}}美国\\ (United States)\\ 0.31\end{tabular} & \begin{tabular}[c]{@{}c@{}}电影\\ (movie)\\ 0.23\end{tabular}   & \begin{tabular}[c]{@{}c@{}}出炉\\ (officially announce)\\ 0.09\end{tabular}  & \begin{tabular}[c]{@{}c@{}}名单\\ (List)\\ 0.12\end{tabular}  & \begin{tabular}[c]{@{}c@{}}电视\\ (TV)\\ 0.02\end{tabular} & \begin{tabular}[c]{@{}c@{}}时间\\ (time)\\ 0.00\end{tabular}  & 0.54       \\ \hline
\end{tabular}
\label{words_scores}
\end{table*}
\end{CJK*}

Interestingly, our method also outperforms the DeepFM and its variation xDeepFM, \textit{e.g.}, 1.02 $vs.$ 0.09 in terms of MSE. Theoretically, the difference between DeepFM and our method is that: rather than extracting feature by the Multiple Layer Perception (MLP) in DeepFM, our method instead utilizes GNN as feature extractors. The results indicate that GNN as a backbone explores the sparse relationship between key words for predicting popularity of events from web; while MLP is suitable for the dense correlation characteristic in the CTR problem.

Another important observation is that SMN-GAT significantly surpassed the counterpart SMN-GCN. The explanation are two aspects: 1) this might attribute to the dynamic modeling of attention in GAT enables more effective feature representation, and 2) the local attention mechanism in GAT is more suitable to model the diversity of key words than GCN.

\subsubsection{Results on HEP-PH dataset}

Table~\ref{result_hepph} details the performance of the SOTA methods and our method with different GNN
backbones on the HEP-PH dataset. Similar to the observations on the Hot Event dataset, our models yields consistent gains against the SOTA methods over all the evaluation metrics except for OL@30. For instance,  SMN-GAT improves by 3.11, 0.9, 0.05, and 0.05 in  MSE, OL@10, mAP, and NDCG compared to xDeepFM, respectively. This generally highlights the generalizability of our methods in the challenging academic paper popularity modeling.

As shown in Table~\ref{result_hepph}, the results of FNN are unexpectedly better than our method over OL@30. This result may attribute to two aspects: 1) the self excitation and the mutual excitation in SMN are sensitive to the diverse events; therefore, our method achieved the best performances over OL@10 and OL@20 instead of OL@30. 2) the popularity scores of research articles in the HEP-PH dataset essentially are topic-sensitive; consequently, the base excitation in SMN may be too weak to the latent topics for research articles.      

In particular, we have noticed that SMN-GAT did not always surpassed the counterpart SMN-GCN. For instance, SMN-GAT only achieves 0.67 mAP which is significantly lower than 0.71 mAP of SMN-GCN as discussed in Section~\ref{sec:sub:ablation}.

\subsection{Interpretability}

 Table \ref{words_scores} shows the interpretability of the results of SMN by the self excitation scores of key words for various events. Concretely, the table reveals that the key words with the high scores are the highlights that characterize the interestingness of an event. For instance, ``second-hand'' in \# 1 event and ``Golden Globe Award'' in the \# 4 event. On the contrary, key words are just essential for the structure of language, \textit{e.g.}, ``times'' and ``yuan'', have zero scores. These examples underscore the effectiveness of the proposed interpretability via sparsity.

%Additionally, a single keyword can significantly enhance an event's popularity, emphasizing the need to filter keywords based on their popularity for accurate predictions. The scores also contribute to the interpretability of mutual excitation. In Event 4, both wards ``Golden Globe Award'' and ``award'' have high scores, possibly due to mutual excitation between them. We find that these words have a strong connection in collectively representing this event. Overall, the predicted scores can somewhat explain the role of words in the text and the relationships between them.

\begin{table}[t!]
  \centering
    \caption{The comparisons between the STOA methods and our method on the HEP-PH dataset. The best accuracy is in bold, and the second best is underlined.}
  \label{result_hepph}
  \resizebox{\linewidth}{!}{
%\scalebox{0.75}{
    \begin{tabular}{c|c|ccc|cc}
        \hline
        Model& MSE $\downarrow$& OL@10 $\downarrow$& OL@20$\downarrow$& OL@30$\downarrow$& mAP $\uparrow$& NDCG $\uparrow$\\
        \hline\hline
        FM& 6.30& 4.87& 9.06& 12.26& 0.53& 0.65\\\hline
        FFM& 5.72& 4.49& 9.10& 11.47& 0.58& \underline{0.68}\\
        FNN& 27.27& 4.78& \underline{6.45}& \textbf{9.44}& 0.54& 0.63\\
        DeepFM& 3.35& 2.88& 7.24& \underline{11.23}& \underline{0.70}& 0.67\\
        xDeepFM& 3.34& 2.95& 8.03& 11.56& 0.62& \underline{0.68}\\ \hline
        $\text{SMN-GCN}$& \underline{0.94}& \underline{2.47}& 7.90& 12.57& \textbf{0.71}& \textbf{0.73}\\
        $\text{SMN-GAT}$& \textbf{0.23}&\textbf{2.05}& \textbf{6.41}& 12.11& 0.67& \textbf{0.73}\\
        \hline
    \end{tabular}}
\end{table}

\section{Conclusion}

In this paper, we have described a method to model the popularity of events on Web, leading to the results significantly outperforming the SOTA methods.
More importantly, the proposed SMN fuses the one-order and the high-order correlations among the text modality, and later fuses them with the image modality features to extract an excellent feature representation.
There are significant distinctions between the proposed method and the previous studies as follows:
\begin{itemize}
\item To our best knowledge, we firstly propose the novel modules of the self-excitation and mutual-excitation to model keyword popularity scores.    
\item We propose an interpretable deep factorization method to predict web event popularity scores by extracting sparse, significant keywords that clarify the reasons why a key word has a high score.
\end{itemize}
In our experiments, we found that self-excitation and mutual-excitation are crucial for accurately predicting popularity. The proposed SMN achieves outstanding performance on both the Hot events dataset and the HEP-PH dataset.
The promising results of this paper motivate a further examination of the proposed SMN. Firstly, we will explore better modeling approaches for data representation in the image modality. Secondly, we will investigate how to achieve more effective interaction between the image and text modalities.

%\bibliography{IEEEabrv,refs}
%\bibliographystyle{IEEEabrv}
\bibliographystyle{IEEEtran}
\bibliography{refs}

\begin{thebibliography}{10}
\providecommand{\url}[1]{#1}
\csname url@samestyle\endcsname
\providecommand{\newblock}{\relax}
\providecommand{\bibinfo}[2]{#2}
\providecommand{\BIBentrySTDinterwordspacing}{\spaceskip=0pt\relax}
\providecommand{\BIBentryALTinterwordstretchfactor}{4}
\providecommand{\BIBentryALTinterwordspacing}{\spaceskip=\fontdimen2\font plus
\BIBentryALTinterwordstretchfactor\fontdimen3\font minus
  \fontdimen4\font\relax}
\providecommand{\BIBforeignlanguage}[2]{{%
\expandafter\ifx\csname l@#1\endcsname\relax
\typeout{** WARNING: IEEEtran.bst: No hyphenation pattern has been}%
\typeout{** loaded for the language `#1'. Using the pattern for}%
\typeout{** the default language instead.}%
\else
\language=\csname l@#1\endcsname
\fi
#2}}
\providecommand{\BIBdecl}{\relax}
\BIBdecl

\bibitem{wang2018learning}
W.~Wang, W.~Zhang, J.~Wang, J.~Yan, and H.~Zha, ``Learning sequential
  correlation for user generated textual content popularity prediction.'' in
  \emph{IJCAI}, 2018, pp. 1625--1631.

\bibitem{manning1999foundations}
C.~Manning and H.~Schutze, \emph{Foundations of statistical natural language
  processing}.\hskip 1em plus 0.5em minus 0.4em\relax MIT press, 1999.

\bibitem{joo2017impact}
T.-M. Joo and C.-E. Teng, ``Impact of social media (facebook) on human
  communication and relationships: A view on behavioral change and social
  unity,'' \emph{International Journal of Knowledge Content Development \&
  Technology}, vol.~7, no.~4, 2017.

\bibitem{figueiredo2011tube}
F.~Figueiredo, F.~Benevenuto, and J.~M. Almeida, ``The tube over time:
  characterizing popularity growth of youtube videos,'' in \emph{Proceedings of
  the fourth ACM international conference on Web search and data mining}, 2011,
  pp. 745--754.

\bibitem{van2008persuasiveness}
G.~Van~Noort, P.~Kerkhof, and B.~M. Fennis, ``The persuasiveness of online
  safety cues: The impact of prevention focus compatibility of web content on
  consumers’ risk perceptions, attitudes, and intentions,'' \emph{Journal of
  Interactive Marketing}, vol.~22, no.~4, pp. 58--72, 2008.

\bibitem{martin2016exploring}
T.~Martin, J.~M. Hofman, A.~Sharma, A.~Anderson, and D.~J. Watts, ``Exploring
  limits to prediction in complex social systems,'' in \emph{Proceedings of the
  25th international conference on world wide web}, 2016, pp. 683--694.

\bibitem{duan2023multi}
S.~Duan, L.~Zhu, and H.~Mo, ``Multi-scale sliding-window fitting model for
  evolution of information dissemination in weibo,'' 2023.

\bibitem{li2022grounded}
L.~H. Li, P.~Zhang, H.~Zhang, J.~Yang, C.~Li, Y.~Zhong, L.~Wang, L.~Yuan,
  L.~Zhang, J.-N. Hwang \emph{et~al.}, ``Grounded language-image
  pre-training,'' in \emph{Proceedings of the IEEE/CVF Conference on Computer
  Vision and Pattern Recognition}, 2022, pp. 10\,965--10\,975.

\bibitem{kang2022forget}
H.~Kang, R.~J.~L. Mina, S.~R.~H. Madjid, J.~Yoon, M.~Hasegawa-Johnson, S.~J.
  Hwang, and C.~D. Yoo, ``Forget-free continual learning with winning
  subnetworks,'' in \emph{International Conference on Machine Learning}.\hskip
  1em plus 0.5em minus 0.4em\relax PMLR, 2022, pp. 10\,734--10\,750.

\bibitem{radford2021learning}
A.~Radford, J.~W. Kim, C.~Hallacy, A.~Ramesh, G.~Goh, S.~Agarwal, G.~Sastry,
  A.~Askell, P.~Mishkin, J.~Clark \emph{et~al.}, ``Learning transferable visual
  models from natural language supervision,'' in \emph{International conference
  on machine learning}.\hskip 1em plus 0.5em minus 0.4em\relax PMLR, 2021, pp.
  8748--8763.

\bibitem{pinto2013using}
H.~Pinto, J.~M. Almeida, and M.~A. Gon{\c{c}}alves, ``Using early view patterns
  to predict the popularity of youtube videos,'' in \emph{Proceedings of the
  sixth ACM international conference on Web search and data mining}, 2013, pp.
  365--374.

\bibitem{wang2018factorization}
W.~Wang, W.~Zhang, and J.~Wang, ``Factorization meets memory network: Learning
  to predict activity popularity,'' in \emph{Database Systems for Advanced
  Applications: 23rd International Conference, DASFAA 2018, Gold Coast, QLD,
  Australia, May 21-24, 2018, Proceedings, Part II 23}.\hskip 1em plus 0.5em
  minus 0.4em\relax Springer, 2018, pp. 509--525.

\bibitem{cui2011should}
P.~Cui, F.~Wang, S.~Liu, M.~Ou, S.~Yang, and L.~Sun, ``Who should share what?
  item-level social influence prediction for users and posts ranking,'' in
  \emph{Proceedings of the 34th international ACM SIGIR conference on Research
  and development in Information Retrieval}, 2011, pp. 185--194.

\bibitem{xiao2016modeling}
S.~Xiao, J.~Yan, C.~Li, B.~Jin, X.~Wang, X.~Yang, S.~M. Chu, and H.~Zha, ``On
  modeling and predicting individual paper citation count over time.'' in
  \emph{Ijcai}, 2016, pp. 2676--2682.

\bibitem{zhang2018user}
W.~Zhang, W.~Wang, J.~Wang, and H.~Zha, ``User-guided hierarchical attention
  network for multi-modal social image popularity prediction,'' in
  \emph{Proceedings of the 2018 world wide web conference}, 2018, pp.
  1277--1286.

\bibitem{dimitrov2017makes}
D.~Dimitrov, P.~Singer, F.~Lemmerich, and M.~Strohmaier, ``What makes a link
  successful on wikipedia?'' in \emph{Proceedings of the 26th International
  Conference on World Wide Web}, 2017, pp. 917--926.

\bibitem{cao2017deephawkes}
Q.~Cao, H.~Shen, K.~Cen, W.~Ouyang, and X.~Cheng, ``Deephawkes: Bridging the
  gap between prediction and understanding of information cascades,'' in
  \emph{Proceedings of the 2017 ACM on Conference on Information and Knowledge
  Management}, 2017, pp. 1149--1158.

\bibitem{sriram2010short}
B.~Sriram, D.~Fuhry, E.~Demir, H.~Ferhatosmanoglu, and M.~Demirbas, ``Short
  text classification in twitter to improve information filtering,'' in
  \emph{Proceedings of the 33rd international ACM SIGIR conference on Research
  and development in information retrieval}, 2010, pp. 841--842.

\bibitem{zhao2015seismic}
Q.~Zhao, M.~A. Erdogdu, H.~Y. He, A.~Rajaraman, and J.~Leskovec, ``Seismic: A
  self-exciting point process model for predicting tweet popularity,'' in
  \emph{Proceedings of the 21th ACM SIGKDD international conference on
  knowledge discovery and data mining}, 2015, pp. 1513--1522.

\bibitem{yang2016stacked}
Z.~Yang, X.~He, J.~Gao, L.~Deng, and A.~Smola, ``Stacked attention networks for
  image question answering,'' in \emph{Proceedings of the IEEE conference on
  computer vision and pattern recognition}, 2016, pp. 21--29.

\bibitem{chen2016micro}
J.~Chen, X.~Song, L.~Nie, X.~Wang, H.~Zhang, and T.-S. Chua, ``Micro tells
  macro: Predicting the popularity of micro-videos via a transductive model,''
  in \emph{Proceedings of the 24th ACM international conference on Multimedia},
  2016, pp. 898--907.

\bibitem{wu2017sequential}
B.~Wu, W.-H. Cheng, Y.~Zhang, Q.~Huang, J.~Li, and T.~Mei, ``Sequential
  prediction of social media popularity with deep temporal context networks,''
  in \emph{Proceedings of the 26th International Joint Conference on Artificial
  Intelligence}, 2017, pp. 3062--3068.

\bibitem{zhang2017uncovering}
K.~Zhang, ``Uncovering urban dynamics via cross-modal representation
  learning,'' Ph.D. dissertation, University of Illinois at Urbana-Champaign,
  2017.

\bibitem{nojavanasghari2016deep}
B.~Nojavanasghari, D.~Gopinath, J.~Koushik, T.~Baltru{\v{s}}aitis, and L.-P.
  Morency, ``Deep multimodal fusion for persuasiveness prediction,'' in
  \emph{Proceedings of the 18th ACM international conference on multimodal
  interaction}, 2016, pp. 284--288.

\bibitem{van2018learning}
G.~van Tulder and M.~de~Bruijne, ``Learning cross-modality representations from
  multi-modal images,'' \emph{IEEE transactions on medical imaging}, vol.~38,
  no.~2, pp. 638--648, 2018.

\bibitem{tan2019lxmert}
H.~Tan and M.~Bansal, ``Lxmert: Learning cross-modality encoder representations
  from transformers,'' \emph{arXiv preprint arXiv:1908.07490}, 2019.

\bibitem{kang2012deep}
Y.~Kang, S.~Kim, and S.~Choi, ``Deep learning to hash with multiple
  representations,'' in \emph{2012 IEEE 12th International Conference on Data
  Mining}.\hskip 1em plus 0.5em minus 0.4em\relax IEEE, 2012, pp. 930--935.

\bibitem{antol2015vqa}
S.~Antol, A.~Agrawal, J.~Lu, M.~Mitchell, D.~Batra, C.~L. Zitnick, and
  D.~Parikh, ``Vqa: Visual question answering,'' in \emph{Proceedings of the
  IEEE international conference on computer vision}, 2015, pp. 2425--2433.

\bibitem{karpathy2015deep}
A.~Karpathy and L.~Fei-Fei, ``Deep visual-semantic alignments for generating
  image descriptions,'' in \emph{Proceedings of the IEEE conference on computer
  vision and pattern recognition}, 2015, pp. 3128--3137.

\bibitem{keneshloo2016predicting}
Y.~Keneshloo, S.~Wang, E.-H. Han, and N.~Ramakrishnan, ``Predicting the
  popularity of news articles,'' in \emph{Proceedings of the 2016 SIAM
  international conference on data mining}.\hskip 1em plus 0.5em minus
  0.4em\relax SIAM, 2016, pp. 441--449.

\bibitem{lipton2018mythos}
Z.~C. Lipton, ``The mythos of model interpretability: In machine learning, the
  concept of interpretability is both important and slippery.'' \emph{Queue},
  vol.~16, no.~3, pp. 31--57, 2018.

\bibitem{li2022interpretable}
X.~Li, H.~Xiong, X.~Li, X.~Wu, X.~Zhang, J.~Liu, J.~Bian, and D.~Dou,
  ``Interpretable deep learning: Interpretation, interpretability,
  trustworthiness, and beyond,'' \emph{Knowledge and Information Systems},
  vol.~64, no.~12, pp. 3197--3234, 2022.

\bibitem{mnih2014recurrent}
V.~Mnih, N.~Heess, A.~Graves \emph{et~al.}, ``Recurrent models of visual
  attention,'' \emph{Advances in neural information processing systems},
  vol.~27, 2014.

\bibitem{ragno2022prototype}
A.~Ragno, B.~La~Rosa, and R.~Capobianco, ``Prototype-based interpretable graph
  neural networks,'' \emph{IEEE Transactions on Artificial Intelligence},
  vol.~5, no.~4, pp. 1486--1495, 2022.

\bibitem{pang-hu-protptype-t-cybernetics-19}
J.~Pang, A.~Hu, Q.~Huang, Q.~Tian, and B.~Yin, ``Increasing interpretation of
  web topic detection via prototype learning from sparse poisson
  deconvolution,'' \emph{IEEE Transactions on Cybernetics}, vol.~49, no.~3, pp.
  1072--1083, 2019.

\bibitem{church1990word}
K.~Church and P.~Hanks, ``Word association norms, mutual information, and
  lexicography,'' \emph{Computational linguistics}, vol.~16, no.~1, pp. 22--29,
  1990.

\bibitem{mikolov2013efficient}
T.~Mikolov, K.~Chen, G.~Corrado, and J.~Dean, ``Efficient estimation of word
  representations in vector space,'' \emph{arXiv preprint arXiv:1301.3781},
  2013.

\bibitem{kipf2016semi}
T.~N. Kipf and M.~Welling, ``Semi-supervised classification with graph
  convolutional networks,'' \emph{arXiv preprint arXiv:1609.02907}, 2016.

\bibitem{lee2019self}
J.~Lee, I.~Lee, and J.~Kang, ``Self-attention graph pooling,'' in
  \emph{International conference on machine learning}.\hskip 1em plus 0.5em
  minus 0.4em\relax PMLR, 2019, pp. 3734--3743.

\bibitem{pang2018increasing}
J.~Pang, A.~Hu, Q.~Huang, Q.~Tian, and B.~Yin, ``Increasing interpretation of
  web topic detection via prototype learning from sparse poisson
  deconvolution,'' \emph{IEEE Transactions on Cybernetics}, vol.~49, no.~3, pp.
  1072--1083, 2018.

\bibitem{leskovec2005graphs}
J.~Leskovec, J.~Kleinberg, H.~Y. Faloutsos, A.~Rajaraman, and J.~Leskovec,
  ``Graphs over time: densification laws, shrinking diameters and possible
  explanations,'' in \emph{Proceedings of the eleventh ACM SIGKDD international
  conference on Knowledge discovery in data mining}, 2005, pp. 177--187.

\bibitem{velivckovic2017graph}
P.~Veli{\v{c}}kovi{\'c}, G.~Cucurull, A.~Casanova, A.~Romero, P.~Lio, and
  Y.~Bengio, ``Graph attention networks,'' \emph{arXiv preprint
  arXiv:1710.10903}, 2017.

\bibitem{rendle2010factorization}
S.~Rendle, ``Factorization machines,'' in \emph{2010 IEEE International
  conference on data mining}.\hskip 1em plus 0.5em minus 0.4em\relax IEEE,
  2010, pp. 995--1000.

\bibitem{juan2016field}
Y.~Juan, Y.~Zhuang, W.-S. Chin, and C.-J. Lin, ``Field-aware factorization
  machines for ctr prediction,'' in \emph{Proceedings of the 10th ACM
  conference on recommender systems}, 2016, pp. 43--50.

\bibitem{guo2017deepfm}
H.~Guo, R.~Tang, Y.~Ye, Z.~Li, and X.~He, ``Deepfm: a factorization-machine
  based neural network for ctr prediction,'' \emph{arXiv preprint
  arXiv:1703.04247}, 2017.

\bibitem{lian2018xdeepfm}
J.~Lian, X.~Zhou, F.~Zhang, Z.~Chen, X.~Xie, and G.~Sun, ``xdeepfm: Combining
  explicit and implicit feature interactions for recommender systems,'' in
  \emph{Proceedings of the 24th ACM SIGKDD international conference on
  knowledge discovery \& data mining}, 2018, pp. 1754--1763.

\bibitem{zhang2016deep}
W.~Zhang, T.~Du, and J.~Wang, ``Deep learning over multi-field categorical
  data: --a case study on user response prediction,'' in \emph{Advances in
  Information Retrieval: 38th European Conference on IR Research, ECIR 2016,
  Padua, Italy, March 20--23, 2016. Proceedings 38}.\hskip 1em plus 0.5em minus
  0.4em\relax Springer, 2016, pp. 45--57.

\bibitem{loshchilov2016stochastic}
I.~Loshchilov and F.~S. Hutter, ``Stochastic gradient descent with warm
  restarts,'' 2016.

\end{thebibliography}

\vfill

\end{document}